\documentclass[prb,showpaces,preprintnumbers,amsmath,amssymb,twocolumn,superscriptaddress,longbibliography]{revtex4-1}
\usepackage{graphicx}
\usepackage{dcolumn}
\usepackage{bm}
\usepackage{amssymb}
\usepackage{amsmath}
\usepackage{epstopdf}
\usepackage{booktabs}
\usepackage{threeparttable}
\usepackage[pdfstartview=FitH, CJKbookmarks=true, bookmarksnumbered=true, bookmarksopen=true, colorlinks, pdfborder=001, linkcolor=blue, anchorcolor=blue, citecolor=blue]{hyperref}
\begin{document}
\title{Extremely large magnetoresistance and Fermi surface topology of PrSb}
\author{F. Wu}
\affiliation{Center for Correlated Matter and Department of Physics, Zhejiang University, Hangzhou 310058, China}
\author{C. Y. Guo}
\affiliation{Center for Correlated Matter and Department of Physics, Zhejiang University, Hangzhou 310058, China}
\author{M. Smidman}
\affiliation{Center for Correlated Matter and Department of Physics, Zhejiang University, Hangzhou 310058, China}
\author{J. L. Zhang}
\affiliation{Anhui Province Key Laboratory of Condensed Matter Physics at Extreme Conditions, High Magnetic Field Laboratory of the Chinese Academy of Sciences, Hefei 230031, Anhui, China}
\author{H. Q. Yuan}
\email{hqyuan@zju.edu.cn}
\affiliation{Center for Correlated Matter and Department of Physics, Zhejiang University, Hangzhou 310058, China}
\affiliation{Collaborative Innovation Center of Advanced Microstructures, Nanjing 210093, China}

\date{\today}

\begin{abstract}
We report magnetotransport measurements of PrSb in high magnetic fields. Our results show that PrSb exhibits extremely large magnetoresistance(XMR) at low temperatures. Meanwhile angle-dependent magnetoresistance measurements were used to probe the Fermi surface via Shubnikov-de Haas (SdH) oscillations. The angular dependence of the frequencies of the $\alpha$-branch indicate a two-dimensional character for this Fermi surface sheet, while the effective mass of this branch as a function of angle shows a four-fold signature. The evolution of the Fermi surface with field was also studied up to 32~T. An enlargement of the Fermi surface up to 14~T is observed, before the oscillation frequencies become constant at higher fields. Meanwhile our analysis of the residual Landau index from the high field data reveals a zero Berry phase and therefore trivial topology of the Fermi surface.
\begin{description}
\item[PACS number(s)]

\end{description}
\end{abstract}

\maketitle

\section{INTRODUCTION}
Materials with nontrivial electronic band structures have been studied intensively in recent years\cite{Hre,TSCTI,Liu,XDPRX}. For instance, topological insulators are bulk insulators but as a consequence of the topology of the electronic bands, the surface states are gapless\cite{Hre,chen2009experimental}, while in Dirac and Weyl semimetals\cite{Liu,XDPRX,CavaCd3As2,huang2015observation,NbAsLYK,WTe2FDL,hirschberger2016chiral}, the symmetry protected bands have linear crossings with massless excitations. Topologically non-trivial band structures can lead to distinctive transport behaviors, such as a $\pi$ Berry phase from the Landau index analysis\cite{mikitik1999manifestation,zhang2005experimental,murakawa2013detection,huang2015observation}, an extremely large magnetoresistance (XMR)\cite{huang2015observation,shekhar2015extremely,liang2015ultrahigh,ali2014large}, and in Weyl semimetals the chiral anomaly can cause a negative longitudinal magnetoresistance  \cite{CavaCd3As2,huang2015observation,NbAsLYK,WTe2FDL}.

Recently, the observation of XMR in the $X$(Sb,Bi) series of compounds, where $X$ is a rare earth element, has attracted much attention \cite{LaSbCava,DHLaSb,Zeng2015Topological,LaBiPRB,LaXFDL,LaBiIOP,CeSbWeyl,NdSbPRB,NdSbIOP,NdSbHField,Nayak2017Multiple,Cavatwo,LaSbTheory,YSbSr}. In these materials the temperature dependence of the resistivity has a plateau at low temperatures upon applying a magnetic field. In LaSb, it was suggested that this behavior has a similar origin to that of SmB$ _6 $, where it is proposed to be related to the appearance of the surface state\cite{LaSbCava}. However, angle resolved photoemission spectroscopy (ARPES) measurements provide evidence that the band topology in LaSb is trivial\cite{DHLaSb}. Meanwhile for LaBi,  ARPES measurements show multiple Dirac cones near the Fermi level and it is proposed to show a non-trivial  band topology with a $\pi$ Berry phase \cite{LaXFDL,Nayak2017Multiple,LaBiIOP}. In CeSb, a negative longitudinal magnetoresistance (MR) was observed, which provides evidence for the presence of the chiral anomaly in the field-induced ferromagnetic state due to time reversal symmetry breaking\cite{CeSbWeyl}. Consequently  it is of interest to explore how the topology and Fermi surface evolves upon changing the $f$-electron configuration in this series of compounds. In addition, since $X$(Sb,Bi) materials all have moderately high quantum oscillation frequencies, it is important to extend the MR measurements to higher fields, in order to reliably determine the Berry phases.

PrSb has a simple cubic rock-salt structure similar to other $X$(Sb,Bi) materials and  has a paramagnetic ground state without any magnetic transitions \cite{tsuchida1965magnetic,Elastic}. The trivalent Pr$^{3+}$ ions have a $4f^2$ configuration that results in a large local moment at the Pr site\cite{PhysRevLett.39.1028}, which contributes to the large magnetic field induced moments revealed by magnetization measurements\cite{Kido1992De}. Crystal field splitting leads to a singlet ground state and the anisotropic exchange gives rise to singlet-singlet excitations at low temperatures near the $X$ point \cite{NeutronPressure,Elastic}. Previous de Haas-van Alphen effect studies indicate that the extremal cross sectional areas  of the electron orbits become larger with increasing field, resulting in a nonlinear relationship between the number of filled Landau levels and the inverse magnetic field.\cite{Kido1992De,Kido1993De}

Here we present a detailed magnetotransport study on PrSb. An extremely large magnetoresistance which reaches $\rho(H)$/$\rho(0)\approx4300$ at 32~T is observed. To probe the Fermi surface via the Shubnikov-de Haas (SdH) effect, angular dependent MR measurements are reported. The angular dependence of the quantum oscillation frequency corresponding to the $\alpha$-branch shows the signature of a two-dimensional nature of the corresponding Fermi surface, while a four-fold symmetric variation of the effective mass is detected. A detailed analysis of the quantum oscillations indicates an enlargement of the Fermi surface with increasing field, which becomes nearly constant when the field exceeds 14 T.  The analysis of the Landau index of this enlarged Fermi surface shows that at high fields the Fermi surface of PrSb is topologically trivial.

\section{EXPERIMENTAL DETAILS}

Single crystals of PrSb were synthesized using a flux method described in Ref. \onlinecite{P1992Growth}. The mixture of Pr, Sb, Sn elements in a molar ratio of 1:1:10  were slowly cooled from 1100$^\circ$C down to 780$^\circ$C, before centrifuging to remove the Sn flux. The resulting crystals were cubic with a typical size of 3mm$\times$3mm$\times$3mm.

 Resistivity and the angular dependent magnetoresistance measurements were performed using a Quantum Design Physical Property Measurement System (PPMS) from 300 K to 2 K with a maximum applied field of 9~T using a field rotation option. Four Pt wires were attached to the sample by spot welding and the current was applied parallel to the [100] direction. Magnetic susceptibility measurements were performed using both a Quantum Design superconducting quantum interference device (SQUID) magnetometer, and the vibrating sample magnetometer option for the PPMS.  The low temperature magnetoresistance were measured in a He$ ^3 $ system with base temperature of 0.27~K and a maximum applied magnetic field of 15~T. The high field magnetoresistance measurements were performed at the High Magnetic Field Laboratory of the Chinese Academy of Sciences.

\section{RESULTS AND DISCUSSION}

\begin{figure}[t]
\begin{center}
  \includegraphics[width=\columnwidth]{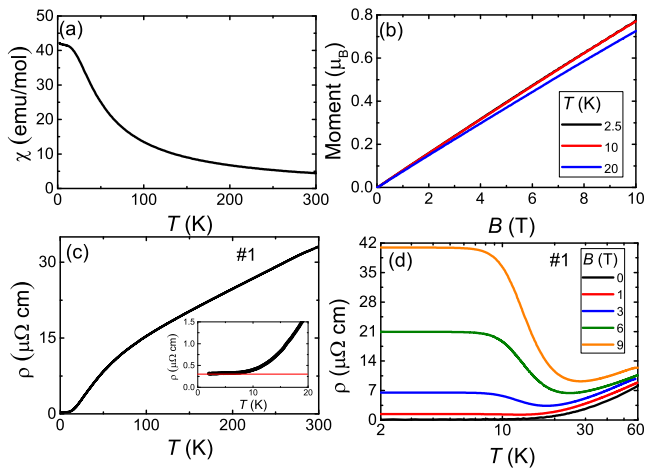}
\end{center}
	\caption{(Color online) (a) Temperature dependence of the magnetic susceptibility of PrSb in an applied field of 0.1 T. (b) Magnetization as a function of applied field at three temperatures, up to a field of 10~T. (c) Temperature dependence of the resistivity  of PrSb sample~\#1 from 300~K to 2 K.The inset displays an enlargement of the low temperature region. (d) Temperature dependence of the resistivity of sample \#1 in several applied magnetic fields. }
   \label{figure1}
\end{figure}

 Figure~\ref{figure1}(a) displays the temperature dependent magnetic susceptibility of PrSb, where there is a broad shoulder at around 10~K, which is consistent with previous reports\cite{monachesi1994optical}. The low temperature behavior can be explained by the crystal field effect commonly observed in Pr-based Van Vleck paramagnets, such as PrPt$_4$Ge$_{12}$\cite{PrPtGe}. No long range magnetic ordering is found down to 2~K and by fitting using the Curie-Weiss law above 50~K, an  effective magnetic moment of $ \mu_{eff} $ = 3.3~$\mu_B$ and a Curie Weiss temperature $ \Theta_P $ = -5 K are obtained. These values are close to the previous study\cite{tsuchida1965magnetic}, and suggest a trivalent nature of the Pr atoms with two fully localized $f$-electrons. The field-dependent magnetization measurements shown in Fig.~\ref{figure1}(b) reveal a nearly linear field dependence of the magnetization at different temperatures below 20~K. At 10~T the magnetization reaches  0.8 $ \mu_B$/Pr, consistent with a magnetic moment of 0.9 $ \mu_B $ at 10~T and 15~K \cite{schoenes1990magneto}. The resistivity of PrSb from 300 K to 2 K is presented in Fig.~\ref{figure1}(c), which shows  typical metallic behavior in zero applied field. The  enlarged view of the low temperature range in the inset shows no anomaly arising from the superconducting transition of Sn, indicating a lack of residual flux in the samples. The residual resistivity is about 0.3~$\mu\Omega$~cm, and the residual resistance ratio is about 110, indicating the high quality of the crystals. Figure~\ref{figure1}(d) displays the temperature dependence of the resistivity when different magnetic fields are applied perpendicular to the current. Unlike in zero field where simple metallic behavior is observed, in the presence of a magnetic field there is an increase of the resistivity upon lowering the temperature, which saturates below 10~K. Similar behavior has also been reported in LaSb\cite{LaSbCava} and YSb\cite{YSbSr}. Although in LaSb this behavior was  reported to be analogous to the topological Kondo insulator SmB$_6$, further studies concluded that this plateau is more likely due to electron-hole compensation\cite{DHLaSb}.

\begin{figure*}
\begin{center}
  \includegraphics[width=2\columnwidth]{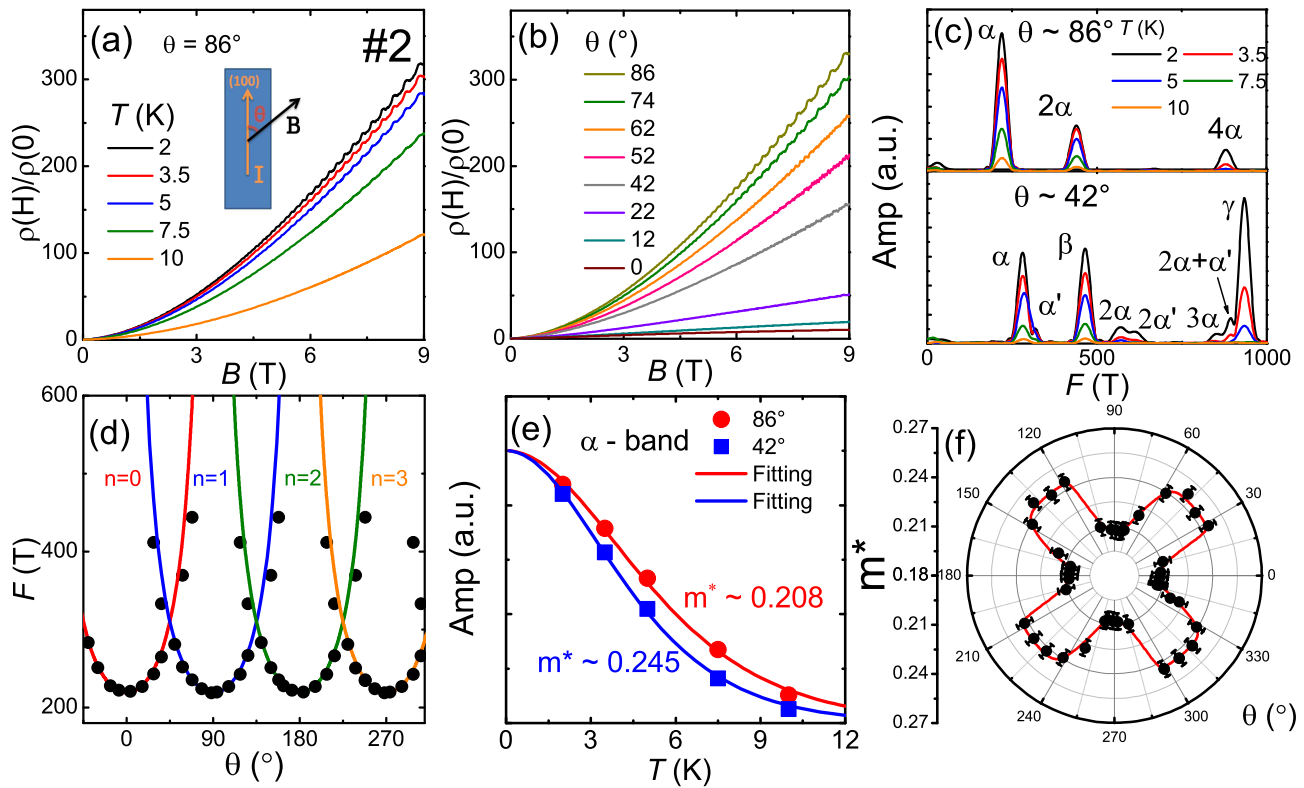}
\end{center}
	\caption{(Color online) (a) Field dependence of the resistivity of sample \#2 at various temperatures with an angle between the magnetic field and current of  $ \theta = 86^\circ $ . The inset illustrates the experimental configuration. (b) Magnetoresistance at 2~K for different $\theta$. (c) The  fast Fourier transform  analysis of the measurements at $\theta$ $\sim$ 86$^\circ$ and 42$^\circ$. (d) Angular dependence of the SdH oscillation frequencies for the $\alpha$-band, where the field is rotated in the (100) plane. (e) Temperature dependence of the quantum oscillation amplitudes of the $\alpha$-band for two values of $\theta$, fitted with the Lifshitz-Kosevich formula. (f) The angular dependence of the effective mass  of the $\alpha$-band obtained from fitting the  Lifshitz-Kosevich formula at different $\theta$.}
   \label{figure2}
\end{figure*}

We also performed magnetoresistance measurements at various temperatures, with fields applied at different angles $\theta$ to the current direction (along [100]), as illustrated in Fig.~\ref{figure2}(a). Figure~\ref{figure2}(a) also displays the MR at $\theta\sim86^\circ$ at several temperature, which shows an XMR effect similar to other $X$Sb compounds, where the MR exceeds  300 at 2 K and 9 T. Upon tilting the magnetic field away from the current direction, the MR decreases  but no negative MR is observed and therefore there is no indication of a chiral anomaly, unlike CeSb where such evidence was found \cite{CeSbWeyl}. Upon rotating the sample, we studied the angular dependence of the oscillation frequency and effective mass. As the frequency of the quantum oscillations is proportional to the extremal cross sectional areas of the Fermi surface, the change of oscillation frequency with angle reflects the Fermi surface structure. The fast Fourier transform (FFT) analysis of the SdH quantum oscillations for two values of $\theta$ are shown in Fig.~\ref{figure2}(c). At $\theta$ $\sim$ 86$^\circ $, the FFT analysis reveals a fundamental frequency of $F_\alpha$ $\sim$ 220~T, and two harmonic frequencies. At $\theta$ $\sim$ $ 42^\circ $, the FFT analysis yields two fundamental $\alpha$ band frequencies at $F_\alpha$ $\sim$ 280~T and $F_{\alpha'}$ $\sim$ 320~T, as well as harmonic frequencies. In addition, there are frequencies associated with the $\beta$ and $\gamma$ bands, $F_\beta$ $\sim$ 463~T and $F_{\gamma}  \sim 934$~T. Note that the $\beta$  frequency can only be clearly observed at certain angles due to the overlap with $F_{2\alpha}$, and we focus on the anisotropy of $\alpha$-band. The oscillation frequencies of the $\alpha$-band as a function of $\theta$ obtained from the FFT  are shown in  Fig.~\ref{figure2}(d). The angular dependence of the oscillation frequencies can be fitted by the expression for a two-dimensional (2D) Fermi surface, $F=F_0/$cos$(\theta-(n\pi/2))$, where $ F_0 \sim 219$~T is the first principal frequency, suggesting that the Fermi surface corresponding to the $\alpha$-band has a two-dimensional nature. Meanwhile the temperature dependence of quantum oscillation amplitudes at two angles are shown in Fig.~\ref{figure2}(e). The solid line shows the result of  fitting with the Lifshitz-Kosevich (LK) formula \cite{shoenberg2009magnetic}, yielding effective masses of m$^*$ of 0.208~m$_0$ for $\theta$ $\sim$ $ 86^\circ $ and 0.245~m$_0$ for $\theta$ $\sim$ $ 42^\circ $. Moreover, the angular dependence of the effective mass from  fitting the LK formula reveals a 4-fold signature, as shown in  Fig.~\ref{figure2}(f). Upon rotating the field, there are eight step-like changes of the effective mass, whereas  between the steps the value stays at a near constant value. Note that the plateau with a higher effective mass around 45$^ \circ $ coincides to where the two bands cross, which leads to an increase of the oscillation frequency. The overall 4-fold signature is consistent with the 4-fold symmetry of the cubic crystal structure in the (100) plane. The abrupt changes of the effective mass  may be related to the structure of the bullet-like Fermi surfaces corresponding to the $\alpha$-band found in other $X$Sb materials \cite{LaBiPRB,YSbSr}, which lie perpendicularly in the Brillouin zone but detailed calculations are required explain this behavior.

\begin{figure}[t]
	\begin{center}
		\includegraphics[width=\columnwidth]{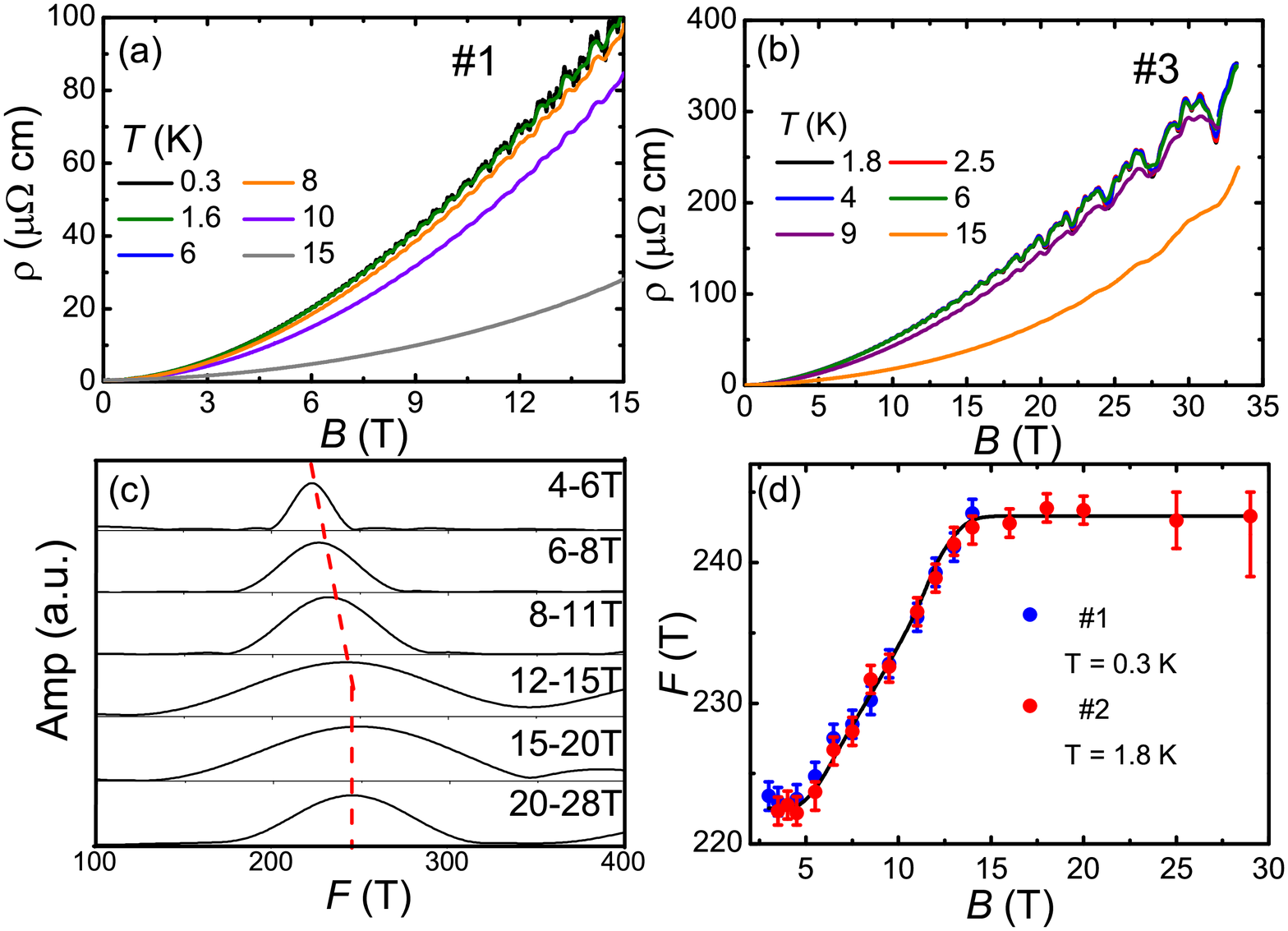}
	\end{center}
	\caption{(Color online) The magnetoresistance at different temperatures is shown up to a maximum field of (a) 15 T and (b) 32 T for samples \#1 and \#3 respectively. (c) FFT of quantum oscillations at 1.8~K in different field ranges showing the position of the fundamental frequency of the $\alpha$-band  $F_\alpha$. (d )  $F_\alpha$ as a function of applied field, for measurements of two samples.}
	\label{figure3}
\end{figure}

To determine whether the large magnetoresistance of PrSb is related to the topological nature of the Fermi surface, we performed high field magnetoresistance measurements on PrSb. Due to the high  frequencies of the quantum oscillations, there can be a reasonably large uncertainty in the extrapolated residual Landau index if the measurements are not performed to sufficiently large fields (small $1/B$). Therefore we measured the SdH oscillations in high magnetic fields to accurately determine the Berry phase. Figure~\ref{figure3}(a) displays the transverse magnetoresistance measured at low temperature down to 0.3~K up to 15~T, while Fig.~\ref{figure3}(b) shows the transverse magnetoresistance measured up to 32~T. The current is along the [100] direction while the magnetic field is perpendicular to the current direction. XMR is observed with clear quantum oscillations, where $\rho(H)$/$\rho(0)$ is about 4300 at $B$ = 32~T. The field  dependent resistivity shows  a quadratic dependence, suggesting that electron hole compensation is the origin of the XMR\cite{pippard1989magnetoresistance}. Figure~\ref{figure3}(c) shows the FFT analysis of the SdH oscillation frequencies of sample \#3 in different field ranges at 1.8~K. It can be seen that in the lower field ranges, the oscillation frequency of the $\alpha$-band increases with increasing magnetic field. The field dependence of these frequencies is displayed in Fig.~\ref{figure3}(d) for both samples, where the field is taken as the midpoint of the field ranges of the FFT. While an increase of the frequency with field was previously reported from a comparison between two field ranges \cite{Kido1992De,Kido1993De}, here we report the detailed evolution. Our results show that there  is an increase of the frequency with field up to around 14~T, while at larger fields the frequency is nearly unchanged. This indicates that at low fields  there is an enlargement of the Fermi surface of PrSb, which becomes near constant at higher fields.

\begin{figure}[t]
	\begin{center}
		\includegraphics[width=\columnwidth]{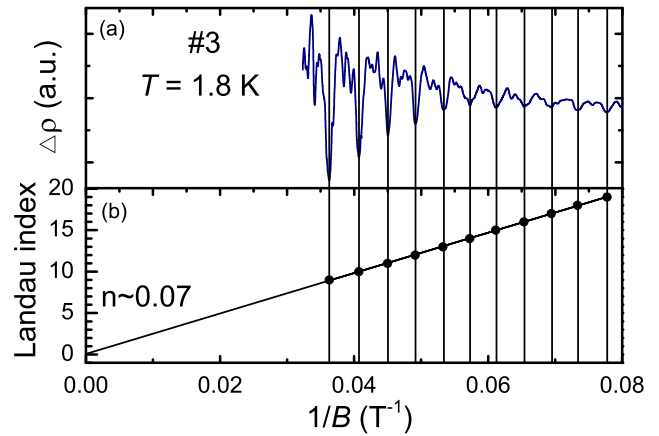}
	\end{center}
	\caption{(Color online) (a) SdH oscillations of PrSb sample \#3 at 1.8~K as a function of $1/B$, obtained from subtracting the background contribution. (b) The $1/B$ dependence of the Landau index, obtained from the valley positions of the SdH oscillations. The solid line shows a linear fit, which reveals a residual Landau index of $n=0.07$.}
	\label{figure4}
\end{figure}

Since  the oscillation frequency increases with increasing field up to 14 T, the residual value cannot be obtained from a linear fit to the inverse field dependence of the Landau index in this field range. Therefore we analyzed the Landau index at higher fields, where the quantum oscillation frequency remains constant. After subtracting the background contribution,  the data at 1.8~K above 15~T in Fig.~\ref{figure3}(b) were plotted as a function of $1/B$, as displayed in  Fig.~\ref{figure4}(a). Despite the strong harmonic oscillations, the positions of the valleys can be clearly resolved and the positions are marked by the vertical lines. Integer Landau levels were assigned to the valley positions, which are shown in Fig.~\ref{figure4}(b). From fitting the data with a linear relationship, the residual Landau index was determined to be 0.07, which corresponds to a zero Berry phase\cite{mikitik1999manifestation}, and hence the Fermi surface of PrSb in this field range is topologically trivial. Our measurements were measured to sufficiently high fields to measure the ninth Landau index and therefore the uncertainty associated with the fitted residual value is small and the results are reliable.

\section{CONCLUSION}
In summary, we report a detailed angle-dependent magnetotransport study of PrSb where we find XMR with $\rho(H)$/$\rho(0)\approx4300$ at 32~T.  We probed the structure of the Fermi surface via SdH measurements, where  we observe 4-fold symmetry in the angular dependence of the effective mass of the $\alpha$-band, which shows several abrupt changes. In addition, there is an increase of the quantum oscillation frequencies of the $\alpha$-band with magnetic field up to 14 T, indicating an enlargement of the Fermi surface, before reaching a near constant level above 14 T. The analysis of the Landau index at high fields shows that there is zero Berry phase in this field range, and therefore the Fermi surface topology is trivial. Meanwhile  the transverse magnetoresistance  shows a near quadratic field dependence and therefore these results indicate that the XMR in PrSb is more likely due to electron-hole compensation rather than topological protection of the surface state. Whether the band topology remains trivial when no magnetic field is applied remains to be determined, which needs to be checked using ARPES measurements. Furthermore, the origin of the widely observed XMR behavior in this series of compounds is still under debate, and the role of the lanthanide elements in influencing the properties and topology of the band structure needs to be explored further.

\begin{acknowledgments}
We thank Y.~Liu, C.~Cao and X.~Lu for valuable discussions and helpful suggestions. This work was supported by the National Key R\&D Program of China (No.~2017YFA0303100, No.~2016YFA0300202), the National Natural Science Foundation of China (No.~U1632275, No.~11474251) and the Science Challenge Project of China (No.~TZ2016004).

\end{acknowledgments}

\end{document}